\documentclass[referee]{aa}
\usepackage{graphicx}
\usepackage{natbib}
\usepackage{txfonts}

\begin{document}

\title{The Evolution of Carbon and Oxygen in the Bulge and Disk of the Milky Way}

\author {G. Cescutti\inst{1}
\thanks {email to: cescutti@oats.inaf.it}
\and  F. Matteucci\inst{1,2}
\and A. McWilliam\inst{3}
\and C. Chiappini\inst{2,4}}

\institute{Dipartimento di Astronomia, Universit\'a di Trieste, via G.B. Tiepolo 11, 34131 Trieste, Italy  
\and  I.N.A.F. Osservatorio Astronomico di Trieste, via G.B. Tiepolo 11, 34131 Trieste, Italy
\and Observatories of the Carnegie Institution of Washington, 813 Santa Barbara St., Pasadena, CA 91101, USA
\and Observatoire Astronomique de l'Universit\'e de Gen\'eve, 51 Chemin des Mailletes, Sauverny,  CH-1290, Switzerland}

\date{Received xxxx / Accepted xxxx}

\abstract
{The evolution of C and O abundances in the Milky Way can impose strong constraints on stellar 
nucleosynthesis
 and help understanding the formation and evolution of our Galaxy.}
{The aim of this paper is to review the measured C and O abundances in the disk and 
bulge of the Galaxy and compare to predictions of Galactic chemical evolution models.}
{We adopt two successful  chemical evolution models for the bulge and the disk, respectively. 
They assume the same 
nucleosynthesis prescriptions but different histories of star formation.}
{The data show a clear
distinction between the trend of [C/O] in the thick and thin Galactic disks, while the
thick disk and bulge trends are indistinguishable with a large ($>$0.5 dex) increase in the C/O 
ratio in the range from $-$0.1 to $+$0.4 dex for [O/H].  In our models we consider
yields from massive stars with and without the inclusion of metallicity-dependent stellar
winds.  The observed increase in the [C/O] ratio with metallicity in the bulge 
and thick disk lies between the predictions utilizing the mass-loss rates of
Maeder (1992) and those of Meynet \& Maeder (2002).  A model without metallicity-dependent
yields completely fails to match the observations.  Thus, the relative increase in carbon
abundance at high metallicity appears to be due to metallicity-dependent stellar winds in massive stars. 
These results also explain the steep decline of the [O/Fe] ratio with [Fe/H] in
the Galactic bulge, while the [Mg/Fe] ratio is enhanced at all [Fe/H].} 
{We conclude that data and models are consistent with a rapid bulge and thick disk formation timescales, and with
metallicity-dependent yields for C and O. The observed too high  [C/O] ratios at low metallicity in the bulge may
be due to an unaccounted source of carbon; very fast rotating metal poor stars, or  
metal-poor binary systems whose envelopes were stripped by Roche lobe overflow.}


\titlerunning{The chemical evolution of Carbon  and Oxygen}

\maketitle

\authorrunning{}

\section{Introduction}

A basic contradiction exists in the interpretation of [O/Fe] and [Mg/Fe] trends in
the Galactic bulge (McWilliam \& Rich 2004; Fulbright et al. 2007).  The [O/Fe]
ratio declines steeply with increasing [Fe/H], while the [Mg/Fe] ratio is enhanced in
all bulge stars, and only declines slightly by [Fe/H]$\sim$$+$0.5 dex.  An often used
paradigm (e.g. Tinsley 1979, Greggio and Renzini 1983, Matteucci and Greggio 1986) 
to explain the alpha-element trend with metallicity, asserts 
that the [O/Fe] ratio declines with [Fe/H] due to the addition of Fe from Type~Ia 
supernovae (SNe) on long timescales.  Since Mg and O are thought to be produced in similar
mass (the most massive) progenitors of Type~II SNe (e.g. Woosley \& Weaver 1995;
henceforth WW95), it is expected that [O/Fe] and [Mg/Fe] should exhibit the same
trend with [Fe/H].  Thus, within this alpha-element paradigm the [O/Fe] trend suggests a
longer formation timescale for the bulge than does the [Mg/Fe] ratio.   McWilliam \& Rich (2004)
suggested that the steep decline in [O/Fe] was actually due to metallicity-dependent winds,
and related to the Wolf-Rayet phenomenon, rather than the addition of Fe from Type~Ia SNe.

In a previous paper (McWilliam  et al. 2008) we analysed the trend
of the [O/Mg] ratio in Galactic disk and bulge stars.  
The observational data indicated that the [O/Mg] showed the same decline with increasing [Mg/H]
in both the bulge and disk, despite the very different formation timescales of these two systems,
and qualitatively consistent with metallicity-dependent O yields.  In
the McWilliam et al. (2008) paper we included 
yields for oxygen from massive stars with metallicity-dependent winds
in detailed chemical evolution models for the bulge and
disk of the Milky Way.  In particular, we adopted the Maeder (1992) yields for 
massive stars including mass loss; these yields are a strong function of the 
stellar mass for stars at solar metallicity and above. For masses larger
than 25$M_{\odot}$, the oxygen production is strongly depressed as a consequence
of this mass loss.  In the presence of a strong mass loss during the Main Sequence phase
of massive stars, a larger amount of He and C are lost relative to stars evolving
at constant mass. This fact produces two main effects: i) a larger production of
He and C from stars with high metal content, ii) a smaller production of oxygen,
due to lack of C which is lost instead of being transformed into O.
In Maeder(1992) and Meynet \& Maeder (2002) this yield effect is important only
for the Z=0.02 models, while the yields of C and O from massive stars
with metallicities of Z=0.004, and below, do not differ substantially from the
yields computed without mass loss (e.g. Woosley \& Weaver 1995). This is due to the
fact that mass
loss in massive stars is a strong function of metallicity.  In McWilliam et al. (2008)
we showed that
the trends of [O/Mg], [Mg/Fe] and [O/Fe] in Galactic bulge and disk stars can be well
reproduced by models with the metallicity-dependent yields of Maeder (1992).
If these metallicity-dependent yields from massive stars are correct, then
the expectation is that the oxygen decline to higher metallicity
stars in the bulge, as evidenced by [O/Mg] and [O/Fe] trends, should be accompanied
by an increase in the [C/O] ratio.  Thus, the [C/O] ratio in the bulge provides a
test for the conclusions of McWilliam et al. (2008).  We note that since C and O are
thought to be produced only during hydrostatic phases of stellar evolution the 
[C/O] versus [O/H] trend in the Galaxy should be considerably simpler to predict
than [C/Fe] and [O/Fe], because iron is produced in the hard to model explosions of Type~Ia
and Type~II SNe in quantities that are not well predicted.  Indeed, even the theoretical
yields for Mg from Type~II SNe (e.g. from WW95) are not well predicted 
and it has been necessary for modellers to employ semi-empirical yields
from Galactic chemical evolution models (e.g. Fran\c cois et al. 2004).

In this paper we analyse the effects of stellar yields with mass loss on carbon and
C/O ratios, both in the bulge and the disk.  As mentioned before, the
decreased O production results in an increased C production and this effect needs
to be tested with the observational data. Besides the Maeder (1992) yields,  we have included in the chemical
evolution models C and O yields from massive stars by Meynet \& Maeder (2002), 
computed with a lower rate of mass loss but including rotation; yields for Z=0.02
stars, which were not given in the original paper, were taken from Chiappini et al. (2003b).
Yields for super-solar metallicity stars are not available.

The paper is organized as follows: in Section 2 we describe the observational data,
in Section 3 we recall the main assumptions of the adopted chemical evolution models,
in Section 4 we describe the adopted stellar yields, in Section 5 we show the 
results for the bulge and the disk of the Milky Way. Finally, in Section 6 some 
conclusions are drawn.

\section{Observational data}\label{data}

We wish to compare the observed trends of [C/O] with [O/H]
in the Galactic thin and thick disks and bulge with predictions of chemical evolution
models including yields from stars with metallicity-dependent winds.  For the
relatively hot unevolved turn-off and main sequence stars carbon and oxygen abundances
are most reliably determined from the forbidden [O~I] and [C~I] lines at 
6300\AA\ and 8727\AA , respectively, because these lines have negligible non-LTE
abundance corrections.  This is unlike the allowed C~I and O~I transitions with large
predicted non-LTE corrections, the estimated values of which have varied significantly
(e.g. Fabbian et al. 2006, 2009).  The insensitivity of the forbidden lines to non-LTE
effects is partly due to the long lifetimes of the upper levels, which allows collisions
to set the populations to their local thermodynamic values.  However, a cause for care
and concern is that the forbidden [C~I] and [O~I] lines are rather weak in the turnoff
and dwarf stars, with equivalent widths typically in the range 2--12 m\AA .  Also, only
in the last few years it has been appreciated (Allende Prieto et al. 2001) that the [O~I] 
line at 6300\AA\ is blended with a Ni~I line; and that the [C~I] feature contains a weak
Fe~I blend (e.g. Bensby \& Feltzing 2006) and lies in the wing of a stronger Si~I line.  
We note, in passing, that the largely ignored [O~I] line at 6363\AA\ is much weaker than the line
at 6300\AA\ and often gives discordant results, which we presume is related to the fact
that it lies in the middle of a broad Ca~I auto-ionization feature which
depresses the central [O~I] line depth; most studies fail to take the Ca~I auto-ionization 
feature into account, resulting in spurious oxygen abundance results.

In this paper we employ forbidden line carbon and oxygen abundances of thin and thick disk
dwarf/turnoff stars from two main sources: Bensby \& Feltzing (2006) and the corrected
oxygen abundances of Nissen \& Edvardsson (1992) and carbon abundances from Gustafsson
et al. (1999) and Andersson \& Edvardsson (1994).  Nissen (private communication) kindly
supplied the revised oxygen abundances, from the [O~I] 6300\AA\ line, corrected for the
Ni~I blend (which was not a known problem in 1992).  While Nissen did not publish the 
corrected oxygen abundance values they do appear in Fig.~9 of Nissen et al. (2002).
We have performed calculations which roughly confirm Nissen's corrections.
In Nissen's private communication to us he also included abundances for two extra, unpublished,
halo stars.   

\begin{figure}[ht!]
\begin{center}
\includegraphics[width=0.99\textwidth]{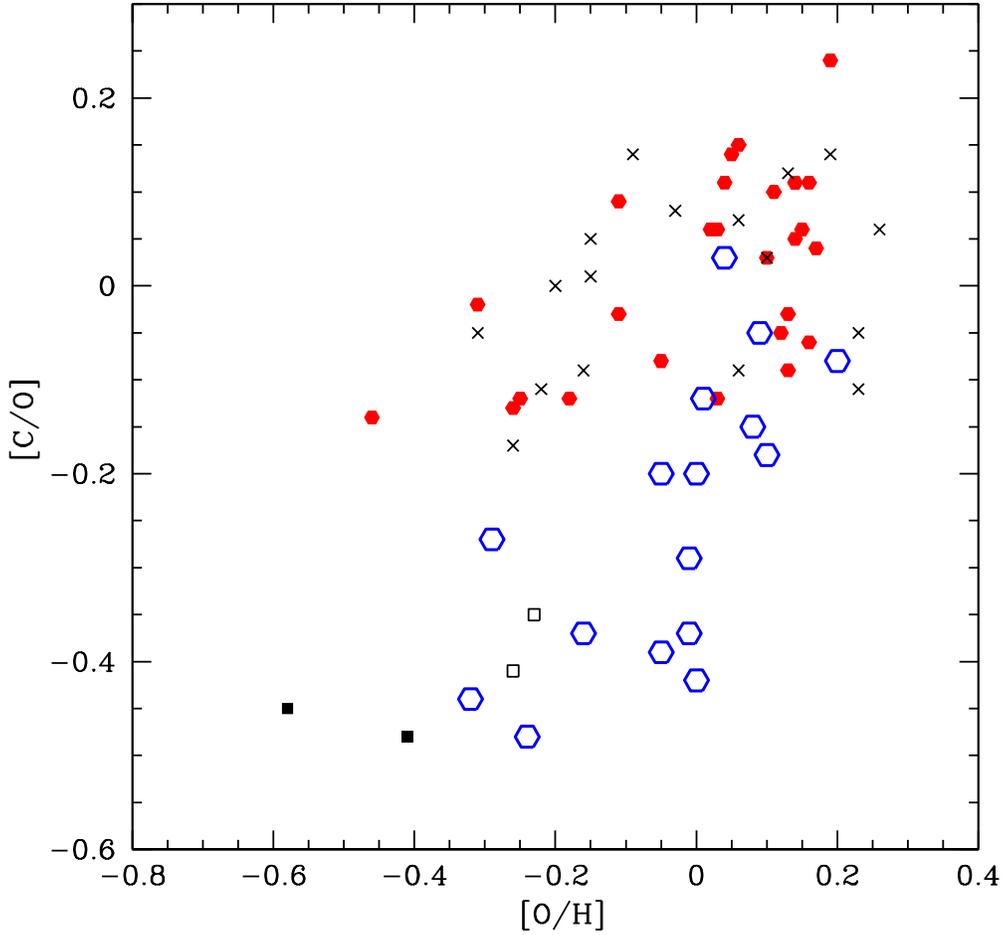}
\caption{[C/O] ratio versus [O/H] for dwarf and turnoff stars in the solar vicinity,
based on forbidden [O~I] and [C~I] lines, which are insensitive to non-LTE effects.
The thin disk stars from corrected Nissen \& Edvardsson (1992), Gustafsson et al. (1999)
and Andersson \& Edvardsson (1994), indicated by black crosses are in good agreement
with the thin disk results of Bensby \& Feltzing (2006, red filled hexagons).
Thick disk stars from Bensby \& Feltzing (2006) are indicated with blue open hexagons;
two thick disk stars from Nissen shown with open black squares are consistent with
the Bensby \& Feltzing (2006) results.  Two halo stars from Nissen (private communication)
are shown as filled black squares.  This plot shows a clear distinction between
the [C/O] ratios in the thin and thick disks: the thick disk [C/O] ratios are
significantly lower than for the thin disk toward lower [O/H].}\label{coohsolardata}
\end{center}
\end{figure}

In Fig.~\ref{coohsolardata}  
we show literature results of [C/O] versus [O/H] for the solar vicinity, based on 
the forbidden [O~I] and [C~I] lines.  The plot shows a clear distinction between the [C/O] 
ratios in the thin and thick disks; the thick disk is characterised by lower [C/O] values,
which decrease with metallicity, and seem to link to the low [C/O] ratio seen in the two
halo stars.  The good agreement between the Nissen data and those of Bensby \& Feltzing (2006)
is encouraging.  We note that the population identifications are based on statistical
probability, so a small number of misidentifications are possible.  The differences in [C/O]
versus [O/H] for the thick and thin disks show that it was not possible to make the thick disk 
from present day thin disk material (e.g. by dynamical heating of the thin disk), or to make
the lower metallicity thin disk from thick disk material (see also Chiappini 2009 and Chiappini et al. in preparation 
for a discussion on the different evolution between the thin and thick disks).
Of course, carbon and oxygen abundance measurements are also possible for red giant stars.  
An advantage is that the [O~I] 6300\AA\ line is stronger in red giants ($\sim$30--70m\AA )
than dwarf/turnoff stars; the 8727\AA\ [C~I] line is also stronger in red giants
(e.g. 11m\AA\ in Arcturus), but it has not been used much in these stars.  A significant
disadvantage posed by the red giants is that stellar evolution effects, during the first
dredge-up phase, cause their atmospheres to be altered from their original CNO composition. 
In the first dredge-up the atmospheres are contaminated by material that previously experienced
proton-burning by CN cycle core burning (e.g. Iben 1964, 1967; Becker \& Iben 1979).  Typically,
solar neighborhood red giant atmospheres show $\sim$0.2 dex carbon depletions, but 1 dex depletions 
are also found (e.g. Lambert \& Ries 1981).
However, because most of the missing carbon is turned into nitrogen the original carbon abundance can
be estimated from the sum of C$+$N.  This is a useful trick for measuring the primordial carbon
abundance in red giants, and especially important for studying the Galactic bulge C/O trend,
since the bulge dwarfs are too faint for routine high resolution abundance studies\footnote{Recently,
Johnson et al. (2008) have measured C/O in a single lensed bulge dwarf star spectrum, but the oxygen
abundance relied on the highly excited allowed transitions at 7770\AA , which are known to suffer
from non-LTE effects.}; thus, red giant stars provide the main probe to effectively explore the
Galactic bulge C/O trend.

\begin{figure}[ht!]
\begin{center}
\includegraphics[width=0.99\textwidth]{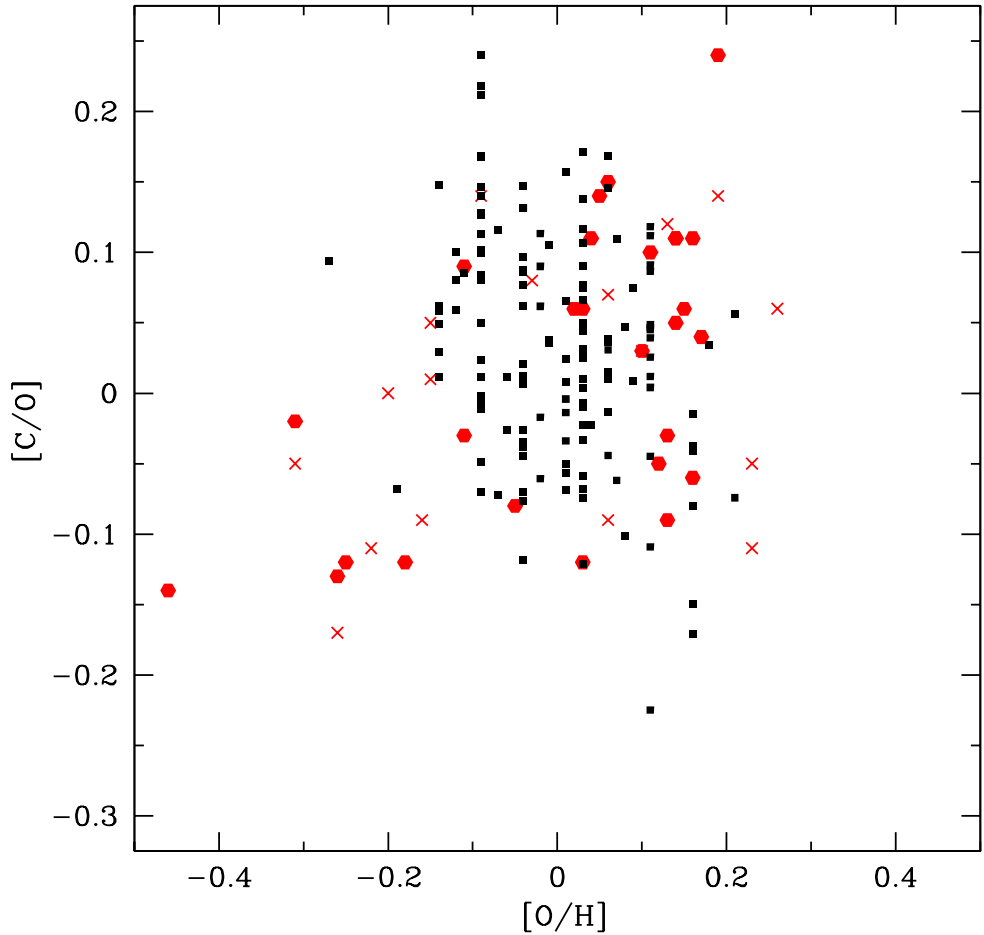}
\caption{[C/O] versus [O/H] for thin disk dwarf/turnoff stars based on forbidden line abundances,
from Bensby \& Feltzing (2006; red filled hexagons) and corrected Nissen (1992; red crosses).
[C/O] versus [O/H] for thin disk red giant stars are from Mishenina et al. (2006; black 
filled squares), but corrected for first dredge-up by adding C$+$N (minus the 
original N based on the solar N/Fe ratio) to estimate the primordial carbon abundance.
Note the good agreement between dwarf/turnoff stars, which are unaffected by first dredge-up, and
the red giant [C/O] ratios from C$+$N - Fe.(N/Fe)$_{\odot}$.  }\label{thindiskdata1}
\end{center}
\end{figure}

In Fig.~\ref{thindiskdata1} 
we show the thin disk [C/O] trend with [O/H]
based on the abundances from dwarf and turnoff stars from Bensby \& Feltzing (2006, red filled
hexagons) and Nissen (private communication, red crosses).  
In addition to these results from dwarf and
turnoff thin disk stars we also include [C/O] ratios based on the [(C$+$N)/O] values in the red
giants of Mishenina et al. (2006; black filled squares); to do this we sum the C and N abundances
but we also subtract an estimate of the original N abundance, based on the solar N/Fe ratio, which
amounts to almost 0.10 dex.  
Fig.~\ref{thindiskdata1} 
shows that the three sources of [C/O] versus [O/H] in the thin disk are 
consistent with a single trend, and a small decrease in [C/O], of about 0.2 dex, from [O/H]=0.0 
to $-$0.4 dex.
 We note that the Mishenina et al. (2006) red giant data lack points at the low [O/H] end
of the diagram; this probably arises from an absence of carbon abundances
for their metal-poor disk stars, due to non-detection of the C$_2$ molecular lines, which they used
as a carbon abundance indicator.  These molecular C$_2$ lines are relatively weak at solar
metallicity, and decrease rapidly in strength to lower metal content.  In addition to the decrease 
in [C/O] at low [O/H] there may be a decrease in [C/O] at the highest metallicity
([O/H] $\ge$$+$0.15 dex), but this conclusion is sensitive to outliers in the plot.
If we use only the current carbon abundances for the Mishenina et al. (2006) red
giants, without the correction for the CN cycle transformation of carbon to nitrogen,
the thin disk red giants are not consistent with the trend established by the two sources
of [C/O] ratios for the thin disk dwarf stars, but instead shows depleted [C/O] values by $\sim$0.2 dex.
We take this difference  as evidence that our practice of combining the C$+$N abundances, 
minus the estimated original N, for red giant stars provides sound results, and should give 
reliable [C/O] ratios when applied to the Galactic bulge red giants. We note, however, that the [C/O] versus [O/H] trend for thin
disk red giant stars in Melendez et al. (2008, henceforth M08) show larger scatter and a
$<$0.1 dex downward shift compared to the dwarf results of Bensby \& Feltzing (2006).  However, the M08 results for the thick disk
are in good agreement with the Bensby \& Feltzing (2006) values, but with somewhat larger scatter.  

%
%

For the bulge [C/O] data we considered abundance results from Fulbright et al. 
(2007), M08, and the oxygen abundances from Zoccali et al. (2006)  with carbon 
results for the same stars given in the companion paper by Lecureur et al. (2007).
Results from the small overlapping studies of Cunha \& Smith (2006) and
Ryde et al. (2009) could also be included in the comparisons.  We omit the abundance
results for bulge M giants from Rich \& Origlia (2005) and Rich et al. (2007) since
they did not measure nitrogen abundances, required to estimate the primordial carbon.

McWilliam, Fulbright and Rich (2009, in progress, henceforth MFR09) have recomputed oxygen
abundances, for stars in the FMR07 study.  In their molecular equilibrium calculations 
they employed the carbon and nitrogen abundances determined from IR spectra in the
same set of stars by M08 (of the 19 bulge stars in M08 18 were studied by FMR07).
The MFR09 abundances are an improvement over FMR07, in part because the FMR07 oxygen
abundances were computed using approximate C/Fe and N/Fe abundance ratios,
expected for typical red giant stars, rather than measured values.  The revised
MFR09 O/Fe ratios were based on line-by-line differential abundances relative to
the sun, using the [O~I] 6300\AA\ line and five transitions of Fe~II.

The differential method employed by MFR09 removed
the $gf$ values from the calculation of the [O/H] and [Fe/H] abundances.  
Furthermore, because formation of the Fe~II lines is similar to the [O~I] lines,
errors in parameters that effect the electron density in the stellar atmospheres 
(e.g. gravity and $\alpha$/Fe ratio) have a negligible effect on the
[O/Fe] ratio determined using these lines.  We note that Zoccali et al.
(2006) did not discuss the use of Fe~II lines or appropriate alpha-enhanced 
model atmospheres in the determination of their [O/Fe] ratios,  because they
used Fe~I lines for the [O/Fe] ratio.  If so, their [O/Fe] ratios would be 
strongly dependent on the gravity, metallicity and $\alpha$/Fe ratios of the input
model atmospheres; indeed, their [C/O] versus [O/H] trend shows a much larger scatter
than the trend found here.  Therefore, we respectfully choose not to employ the
Zoccali et al. (2006) [O/Fe] ratios.
For the [Fe/H] abundances MFR09 retained the FMR07 [Fe/H] values,
which were based on differential Fe~I line abundances relative to the nearby red giant
$\alpha$~Boo, and provide [Fe/H] values with small internal scatter.

We note that the [O/Fe] versus [Fe/H] trend of MFR09 and that of M08
agree very well from $-$0.8$<$[Fe/H]$<$$+$0.5 dex, but below [Fe/H]$\sim$$-$0.8
dex MFR09 [O/Fe] values are larger than M08 by
$\sim$0.1 dex.  It is likely that the two OH lines in the M08 spectra are relatively
weak at these lower metallicities, while the [O~I] lines in FMR07/MFR09 are easily
measured.  Differences between the M08 points and MFR09 values are partly due to the
adopted [Fe/H] abundance; 
we do not know the cause of the difference between FMR07/MFR09 and M08
[O/Fe] trends at low metallicity.
Perhaps the reduced strength of the Fe I lines, or the small number of
detected Fe I lines, in the near-IR spectra of the most metal-poor stars
in M08 resulted in less accurate Fe abundances.  We also note that the
that the optical [O~I] lines in FMR07 are insensitive to non-LTE effects,
while OH lines are sensitive to non-LTE, which is more significant for
metal-poor stars. For these reasons we prefer to employ
the [O/Fe] values computed by MFR09 with the M08 [C/Fe] and [N/Fe] ratios.

As mentioned previously, the C$+$N $-$ Fe.(N/Fe)$_{\odot}$ provides a good
estimate of the original carbon abundance in red giant stars that have experienced
first dredge-up of CN processed material.   For the bulge results we employ 
[O/Fe] from MFR09 and the [C/Fe] and [N/Fe] abundances from M08 to determine
the primordial [C/O] and we use the MFR09 [O/Fe] with FMR07 [Fe/H] ratios to compute 
[O/H].

\begin{figure}[ht!]
\begin{center}
\includegraphics[width=0.99\textwidth]{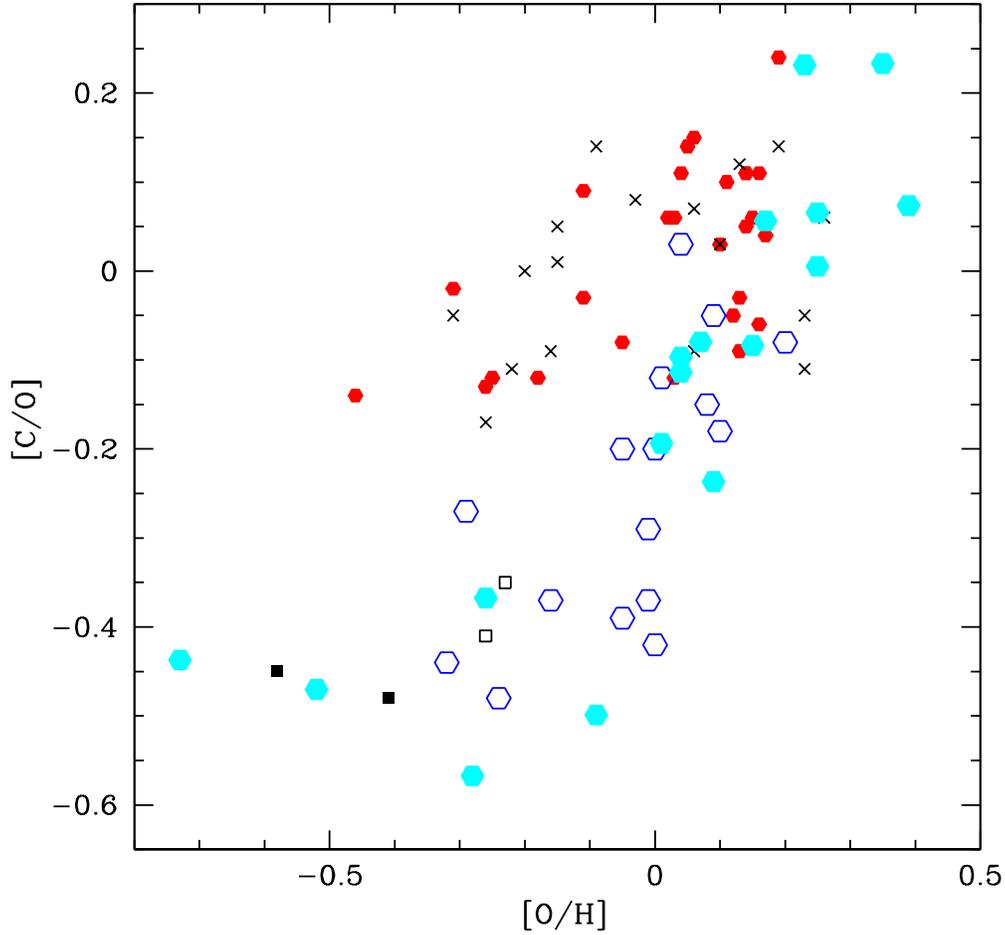}
\caption{ Our estimated primordial [C/O] versus [O/H] trend in Galactic bulge
red giants (cyan filled hexagons), determined from FMR07, MFR09 and M08, compared
with the thin disk (red filled hexagons and black crosses) and thick disk
(blue open hexagons and black open squares) dwarf/turnoff stars (identical to
Fig.~\ref{coohsolardata}).  The primordial values
for the bulge red giants were estimated from {(C$+$N) $-$ Fe.(N/Fe)$_{\odot}$}.  The tight
correlation of [C/O] versus [O/H] in the bulge appears identical to the relationship
seen in the thick disk stars.  }\label{diskbulgedata}
\end{center}
\end{figure}

In Fig.~\ref{diskbulgedata}
we show our adopted bulge trend of primordial [C/O] 
versus [O/H], estimated from the measured carbon and nitrogen abundances of
M08 and the MFR09 [O/FeII] ratios, from the FMR07 EWs.  For
the [O/H] values we add the FMR07 [Fe/H] abundances to the robust [O/FeII] ratios 
of MFR09.  For comparison, the figure also includes the data for thin and thick
disk stars presented in Fig.~\ref{coohsolardata}.
Fig.~\ref{diskbulgedata}
shows a relatively tight correlation of [C/O]
versus [O/H] in the bulge, with a plateau near [C/O]=$-$0.50 dex for metallicities below
[O/H]$\sim$0, but rising, roughly linearly from [O/H]$\sim$$-$0.1, with increasing
metallicity to [C/O]$\sim$$+$0.1--0.2 dex near [O/H]$\sim$$+$0.3 dex.  The most obvious
feature is that the bulge [C/O] versus [O/H] trend is remarkably similar to that of the
thick disk, and quite different than the thin disk.
We note that the 5 stars from Cunha \& Smith (2006) and the 3 stars from Ryde et al. (2009),
most of which are in FMR07 and M08, show similar [C/O] versus [O/H] to those seen in the bulge
in Fig.~\ref{diskbulgedata}, but with larger scatter than our data.  However, there simply 
aren't enough points in either of these two studies for a useful comparison.  Two of the
stars in Ryde et al. (2009) are in Cunha \& Smith (2006), but for one (BW IV-329) their
abundances differ significantly.  In addition, the Ryde et al. (2009) [C/O] and [O/H] values 
for BW~IV-203 place it far from the remaining points in Fig.~\ref{diskbulgedata},
 due to it's enormous nitrogen abundance; also,  FMR07 determined that this star 
was not a member of the bulge by comparing photometric and ionization based gravities, as stated
in that paper.
Given the
much larger sample of stars in MFR09 and M08, the small scatter evident in the
bulge [C/O] ratios of Fig.~\ref{diskbulgedata}, 
the use of the forbidden [O~I] line here, and the uncertain systematics of 
the Ryde et al. (2009) and Cunha \& Smith (2006) results we, respectfully, prefer not to 
use the latter.

\section{The chemical evolution models}

For studying the chemical evolution of the solar neighbourhood, we adopted the model 
of Fran\c cois et al. (2004), which is an implemented version of the original model
by Chiappini et al. (2003a). This model assumes that the Galaxy formed by means 
of two main accretion episodes, one giving rise to the halo and thick disk and 
the other forming the thin disk.
The infalling gas is always assumed to be of primordial composition. Detailed 
nucleosynthesis from low and intermediate mass stars, Type Ia and Type II SNe is
taken into account. The IMF is taken from Scalo (1986).  For details of this model
see Fran\c cois et al. (2004).

For the bulge, we adopted the model of Ballero et al. (2007a), which assumes a rapid formation 
timescale, of 0.3--0.5 Gyr, from gas accumulated during the Halo collapse. 
The efficiency of star formation (star formation per unit mass of gas) is 
20 times 
higher (i.e. $20 Gyr^{-1}$) than in the solar vicinity ($1 Gyr^{-1}$). 
The IMF is flatter than in the solar vicinity, as required by the observed 
bulge stellar metallicity distribution:  we assumed the following
IMF: x=0.95 for stars with $m > 1M_{\odot}$, and 
x=0.33 for stars with $m \le 1M_{\odot}$ (see Ballero et al. 2007a for details of this model).

\section{Nucleosynthesis Prescriptions}{\label{NP}}

In this work we investigate 3 different prescriptions
for the yields from  massive stars, in particular: i)
the metal dependent yields of Woosley \& Weaver (1995, hereafter 
WW95), ii) the metal dependent stellar yields 
from massive stars of Meynet \& Maeder (2002, hereafter MM02), which
take into account the effects of mass loss and rotation on stellar evolution,
used also in Chiappini et al. (2003b),
iii) the yields with mass loss and no rotation of M92.
It is worth noting that both the M92 and MM02 yields from massive stars
include mass loss by stellar winds, but the MM02 calculations adopt a
significantly lower rate of mass loss. The recent mass loss rates are factors
of 2 to 3 times lower than previously adopted by M92 and they include stellar 
rotation.
\\
For low to intermediate mass stars we use in all the models
the van den Hoek \& Groenewegen (1997) metallicity dependent yields
(in particular the tables assuming a mass loss parameter, $\eta$, varying with metallicity).
We also tested the Karakas \& Lattanzio (2007) metallicity dependent yields of C 
and O from low and intermediate mass stars, and found negligible differences in the results 
compared to those obtained  with the van den Hoek \& Groenewegen (1997) yields. We then
 adopted the latter ones for consistency with our previous paper (McWilliam et al. 2008). 

We note that the more recent oxygen yields of Hirschi et al. (2005) are significantly higher 
than MM02, by factors of 2--3.  In fact, Hirschi et al. (2005) stellar evolution calculations 
were carried out with different input parameters (e.g. different amounts of  overshooting).
Moreover they only included material lost in the wind, and nucleosynthesis up to the pre-SN stage;
they did not include any treatment of the supernova explosion, 
explosive nucleosynthesis calculations (e.g. explosive O burning), or other issues related 
to SN nucleosynthesis yields, as discussed in WW95 and many other works.  
Therefore, we avoid combining these different set of yields which otherwise
would produce an artificial metallicity dependency of the oxygen yields.

The M92 and MM02 yields did not perform supernova nucleosynthesis calculations, but they 
employed the Arnett (1991) relationship between O yield and the M$_{\alpha}$ core mass 
to determine their oxygen yields.  The Arnett (1991) O yield relationship with M$_{\alpha}$ 
is very similar to the independent result of WW95.  

For Fe we compute all the models with 
the stellar yields by WW95 at solar metallicity for Type~II SNe, which give the 
best agreement for the solar vicinity (see Fran\c cois et al. 2004).\\
For the yields from SNe Ia we adopt those of Iwamoto et al. (1999) where 
each SN produces 0.6$M_{\odot}$ of Fe.

\section{Results}

We run several models for both the bulge and the solar vicinity with 
yields from massive stars taken from various authors:
A) metallicity-dependent WW95 yields, without mass-loss;\quad
B) WW95 yields for metallicities
below solar and the M92 yields (with mass loss) for $Z \ge Z_{\odot}$;\quad 
C) WW95 yields for metallicities lower than solar and MM02
yields (including mass-loss) for  $Z \ge Z_{\odot}$;\quad 
D) MM02 metal-dependent yields (with mass loss) only.

\begin{figure}[ht!]
\begin{center}
\includegraphics[width=0.99\textwidth]{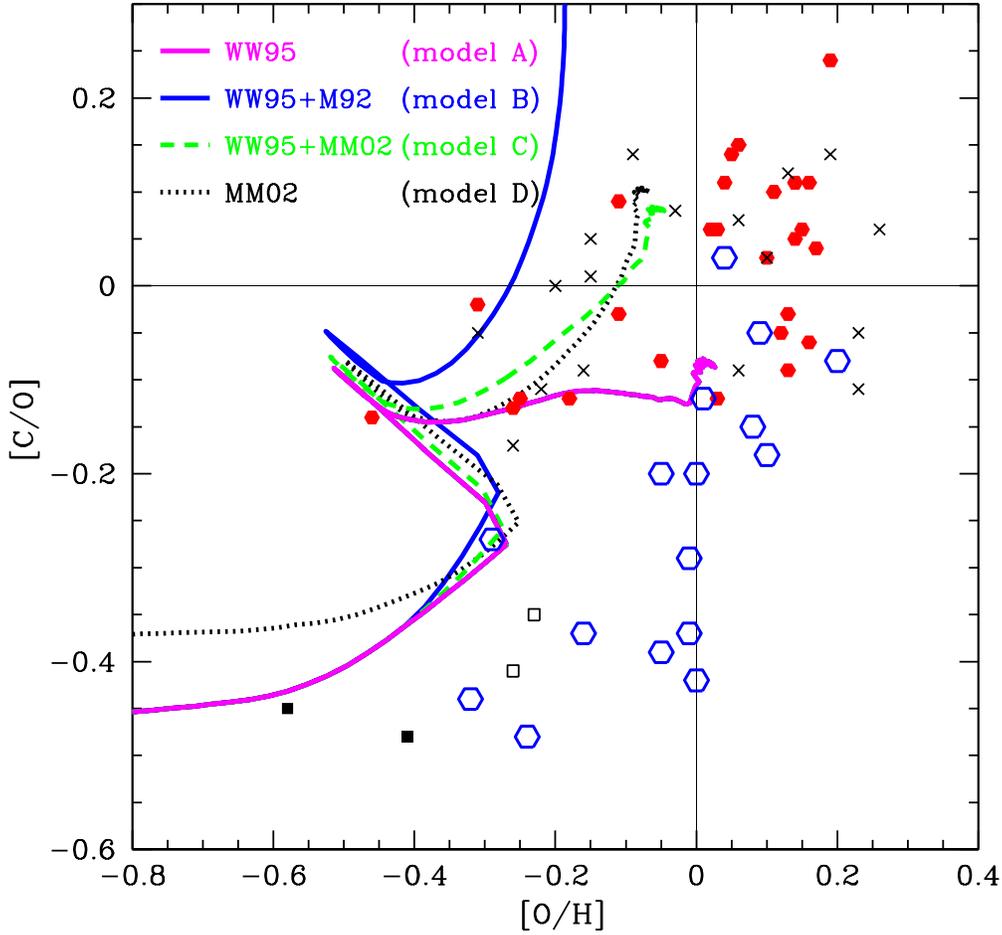}
\caption{Predictions of [C/O] versus [O/H] in the thin disk compared
to measured abundances in the solar vicinity.  See Fig.~\ref{coohsolardata} 
for the key to the data symbols.  We present the results of four different
chemical evolution models, with and without metallicity-dependent yields.  To better
match the [C/O] versus [O/H] slope in the bulge (c.f. Fig.~\ref{bulgetheory}) we interpolate
MM02 and M92 yields between Z=0.004 and Z=0.002.}\label{coohsolartheory}
\end{center}
\end{figure}

In Fig.~\ref{coohsolartheory} we show the results of four chemical evolution
models compared with the solar neighborhood [C/H] versus [O/H] trends for the thin and
thick disks.  A strength of this plot is that only C and O are involved, so complications from
the production of Fe in Type~Ia and Type~II supernovae are avoided.  Because carbon and
oxygen are produced during hydrostatic phases of stellar evolution the yields are much more
tractable than for for Fe.

The predictions for the thin disk [C/O] trend show an increase in the [C/O] ratio associated
with carbon production from relatively low mass stars, on long timescales ($>$1Gyr).  This
partly agrees with Bensby \& Feltzing (2009) conclusion, who argued that the production of
carbon occurred on a similar timescale to the production of Fe, based on a flat [C/Fe]=0.0
trend for both the thin and thick Galactic disks below solar [Fe/H].  Our interpretation
based on the data and our predictions in Fig.~\ref{coohsolartheory}
is that the thin disk shows higher [C/O] ratios than the thick disk, due to increased carbon
yields from low mass stars;
our predictions show such increased carbon yields on long time scales (from low mass stars)
for the thin disk only.  Our thin disk models suggest that approximately 50\% of the
carbon produced along the whole galactic history comes from low mass stars.

We presume that, similar to the bulge, the thick disk evolved more rapidly than the thin disk 
(see also  Melendez et al. 2008), given the similarity of the thick disk and bulge [C/O] versus
[O/H] trend in Fig.~\ref{diskbulgedata}.
Therefore, the thick disk would not be expected to have included the extra carbon produced
on long time scales by low mass stars; this is in contrast to the conclusion of Bensby \& 
Feltzing (2009).

Several features in Fig.~\ref{coohsolartheory} are worthy of note: first, 
the slope of the [C/O] versus [O/H] trend in the thin disk is best matched by models C and D,
which employ the MM02 yields with metal-dependent stellar winds plus rotation.  The higher
mass-loss rates used in the M92 metal-dependent winds result in too steep a [C/O] versus [O/H] 
slope, as if too much carbon is released at the expense of oxygen.  On the other hand model
A, which employed the metal-dependent evolution of WW95 but ignored mass loss from stellar
winds and rotation, gave too shallow a slope in the [C/O] versus [O/H] relation for the
thin disk.  

We draw attention to the inflection points in our thin disk models in Fig.~\ref{coohsolartheory},
where [O/H] appears to decrease temporarily.  This is due to the competition between the star 
formation rate (SFR) and the gas infall rate.
In fact, in the two infall model a gap in the SF is naturally produced at the end
of the halo-thick disk formation.  Such a gap is mainly due to the different infall episodes 
coupled with a threshold in the gas density for SF (see also a discussion of this point in Chiappini et al. 2003b).  
At the inflection
point, star formation is not active because the infalling gas forming the thin disk has still a density below the threshold. 
Therefore, O is no more produced while the thin disk is growing by means of H infalling gas.
At later times, on long timescales, the disk [O/H] increases
again, due to the activation of SF and to the diminishing effect of the declining infall rate.
Also, the [C/O] ratio increases further due to carbon from long-lived stars.

The maximum [O/H] and [C/O] ratios reached by our models is also worth noting:
Fig.~\ref{coohsolartheory}
shows a progression in the maximum [O/H] from the models that
decreases with increasing stellar wind mass-loss rate, while the [C/O] ratio increases
with increasing stellar mass-loss rates (WW95$<$MM02$<$M92).  These maximum ratios are
simply due to the increase in carbon yield, at the expense of oxygen, as the mass-loss
prescription is increased.  Clearly, none of our models reaches the highest
[O/H] values observed in the thin disk.  The solution to this lacuna is not obvious:
perhaps the theoretical nucleosynthesis yields need to be increased (e.g. due
to nuclear reaction rates or stellar evolution effects), or we should include in the models the migration 
of more metal rich stars in the solar vicinity (see Sch\"onrich \& Binney 2009). Ideally, our predictions should match
both the [C/O] trend with [O/H] and the [O/H] distribution function, similar to the metallicity
distribution function (MDF) from [Fe/H] frequently used to constrain the gas infall rate parameter.

It is encouraging that the thin disk [C/O] observations nicely overlap our predictions, at least
for [O/H] less than solar, but zero-point shifts due to the exact value of the solar oxygen and
carbon abundances could slightly diminish this pleasing agreement; Asplund's group have recently
suggested an increased solar oxygen abundance, by 0.05 dex (Scott et al. 2009).  Other
zero-point shifts may occur if theoretical yields are over/under estimated, or sources omitted;
however, differences between predicted trends between the thin disk and bulge/thick will be more robust.

\begin{figure}[ht!]
\begin{center}
\includegraphics[width=0.99\textwidth]{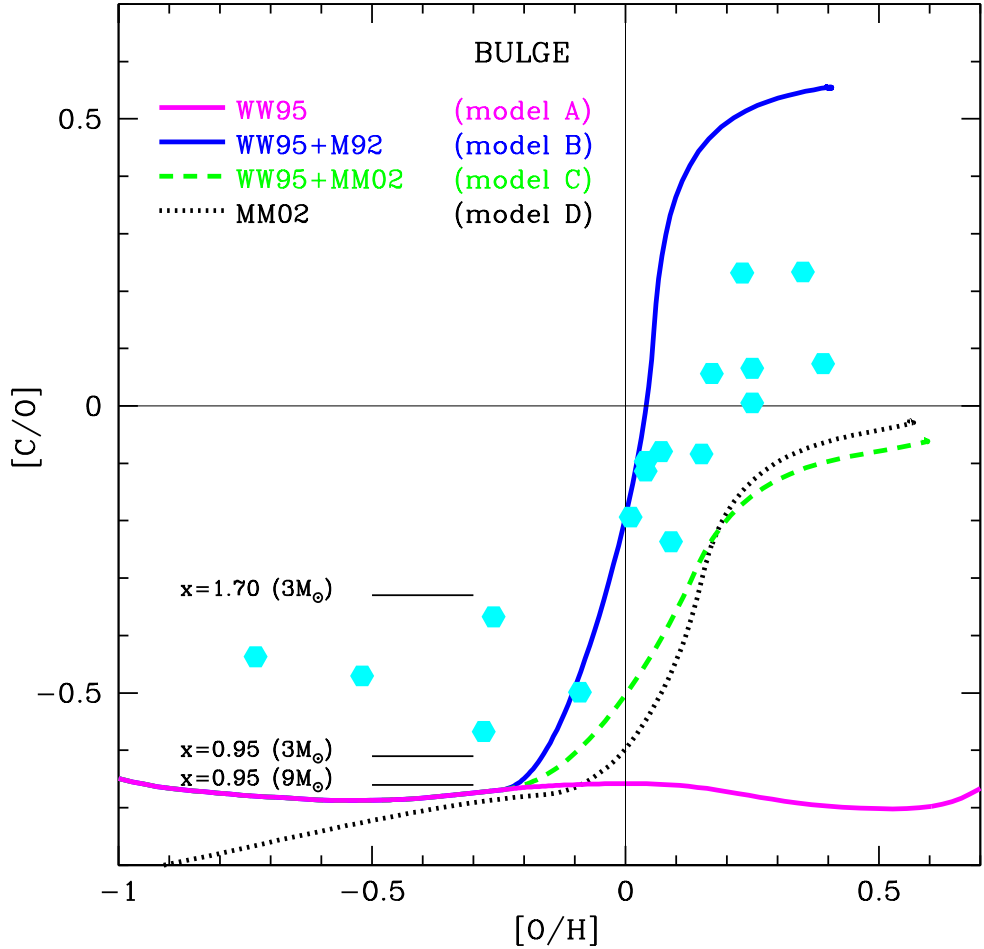}
\caption{Model predictions compared to observed [C/O] versus [O/H] in the Galactic bulge.
Model A (red solid line) shows the expected trend for WW95 yields, which did not include
metallicity-dependent mass-loss.  Model B (blue solid line) shows our predictions with
WW95 yields at low metallicity but M92 yields for solar metal content.  The M92 yields include
the effects of metallicity-dependent winds at relatively high mass-loss rate, but did not
include rotational effects.  Model C (green dashed line) shows the result when the lower
mass-loss rates, but including rotation, of MM02 are included, together with the WW95 yields for low metallicity.  
Model D (black dotted line) employs
only the MM02 yields which include metallicity-dependent winds and rotation.}\label{bulgetheory}
\end{center}
\end{figure}

In Fig.~\ref{bulgetheory}
we show the [C/O] versus [O/H] data and predictions from our
bulge models.  We note that for the models that employ the MM02 and M92 yields we have interpolated
the metallicity-dependent yields between Z=0.02 and Z=0.004.  We believe this to be a reasonable
procedure, but if this interpolation is not adopted then the predicted [C/O] trends rise nearly
vertically, much steeper than the gentle positive slope described by the data; we have employed
this same interpolation of yields over metallicity for the thin disk models previously described.
We first note
that the predictions with the metallicity-dependent yields from WW95 completely fail to
match the bulge observations, while the shape of the observed upward trend in [C/O] with increasing
[O/H] seems to fall between the predictions based on the MM02 and MM92 yields affected by 
metallicity-dependent winds in massive stars.  We take these predictions as strong evidence that
yields from massive stars with metallicity-dependent winds are largely responsible for the increase 
in observed trend of [C/O] and the decrease in observed trends of [O/Mg] (and [O/Fe]) in the bulge.  Thus,
the apparent inconsistency between the enhanced [Mg/Fe] ratios and declining [O/Fe] with [Fe/H]
in the bulge is simply a metallicity effect, as suggested by McWilliam et al. (2008).
Apparently, the metallicity-dependent winds do not decrease the Mg yields as much as the O yields. 
 It is worth noting that
for the thin disk the increase in the [C/O] trend is mainly due to the contribution of intermediate and
low mass stars at low [O/H] because of the low star formation rate in the thin disk, whereas the increase of 
[C/O] trend in the bulge is mainly due to the C produced by massive stars occuring at a high [O/H] because of the fast 
star formation rate in the bulge. In other words, the different star formation rates in the bulge and thin 
disk produce quite different [O/H] - age relationships, with the bulge reaching very fast high [O/H] values.

Closer inspection of Fig.~\ref{bulgetheory} shows that below [O/H]$\sim$$-$0.2 dex the
[C/O] ratio in the bulge is roughly constant, near $-$0.5 dex.  This differs from the
predicted [C/O] level near $-$0.7 dex.  
To explain the difference between the observed and predicted [C/O] ratio in the metal-poor bulge 
one might appeal to the bulge IMF slope for massive stars, because more massive stars have lower [C/O]
yield ratios, due to the larger core masses.  However, the bulge IMF slope has been set by
Ballero et al.  (2007a) at x=0.95 in order to explain the combination of a near-solar mean metallicity
distribution function and the high [Mg/Fe] (which suggests a rapid formation timescale).  Thus the 
bulge IMF, which Ballero et al. (2007a) require to be skewed to more massive stars than the disk, can
only be significantly altered if the bulge did not form rapidly (contrary to the [Mg/Fe] evidence).
The black horizontal lines marked on Fig.~\ref{bulgetheory}
show the [C/O] ratios for two IMF slopes: x=0.95, the Ballero et al. (2007a) bulge value, and x=1.70,
the Kroupa (2002) value for the solar neighborhood (corrected for binaries).  The masses in
parentheses indicate the lower mass limit of the IMF integrations.  
A 9M$_{\odot}$ limit shows the [C/O] ratio produced by Type~II supernovae only, while the 
3M$_{\odot}$ limit indicates the lowest mass star that could have died during the bulge formation 
time ($<$1Gyr).
Thus, to explain the bulge [C/O] plateau at lower [O/H] would require an IMF slope of
approximately x=1.3.

Other possibilities that could increase the predicted [C/O] in the low metallicity bulge
are: 1.  The C production from very fast rotating extremely metal poor stars, but in this case a plateau in [C/O] would 
not be obtained but rather an initial peak followed by  a further decrease (see Chiappini et al. 2006), 
although in a starburst situation, such as in ellipticals or bulges, fast stellar rotators 
could have been present up to larger metallicities, as recently suggested by Pipino et al. (2009).
2. A decrease in the $^{12}$C($\alpha$,$\gamma$)$^{16}$O rate, although we are unable to
give a quantitative estimate or say whether this can reasonably be the case.  3. An extra
source of carbon at low metallicity could be provided by low metallicity massive binaries.  
We recall that an early
suggestion for the production of Wolf-Rayet (WR) stars was through Roche lobe mass-loss/stripping
of the outer envelopes of massive stars (Paczy\'nski 1967).  While this mechanism is not favored as
the main source of WR stars in the Magellanic Clouds and Galaxy, it should have occurred at some
level, even in low metallicity massive binaries.  Given that the stripping of the envelopes of
massive stars by winds at solar metallicities is responsible for the increased carbon and decreased
oxygen yields, it seems possible that metal-poor WR stars formed through Roche-lobe mass loss in
binary systems would also have increased carbon and decreased oxygen yields.  Quantitative 
calculations are required to determine whether this mechanism could have produced the observed
[C/O] ratio in metal-poor bulge stars.  We assume that at high metallicity the Roche lobe stripping
of metal-poor massive binary star envelopes is pre-empted by the effects of metallicity-dependent 
winds, so this particular extra source of carbon probably mainly affects the metal-poor regime.
\begin{figure}[ht!]
\begin{center}
\includegraphics[width=0.99\textwidth]{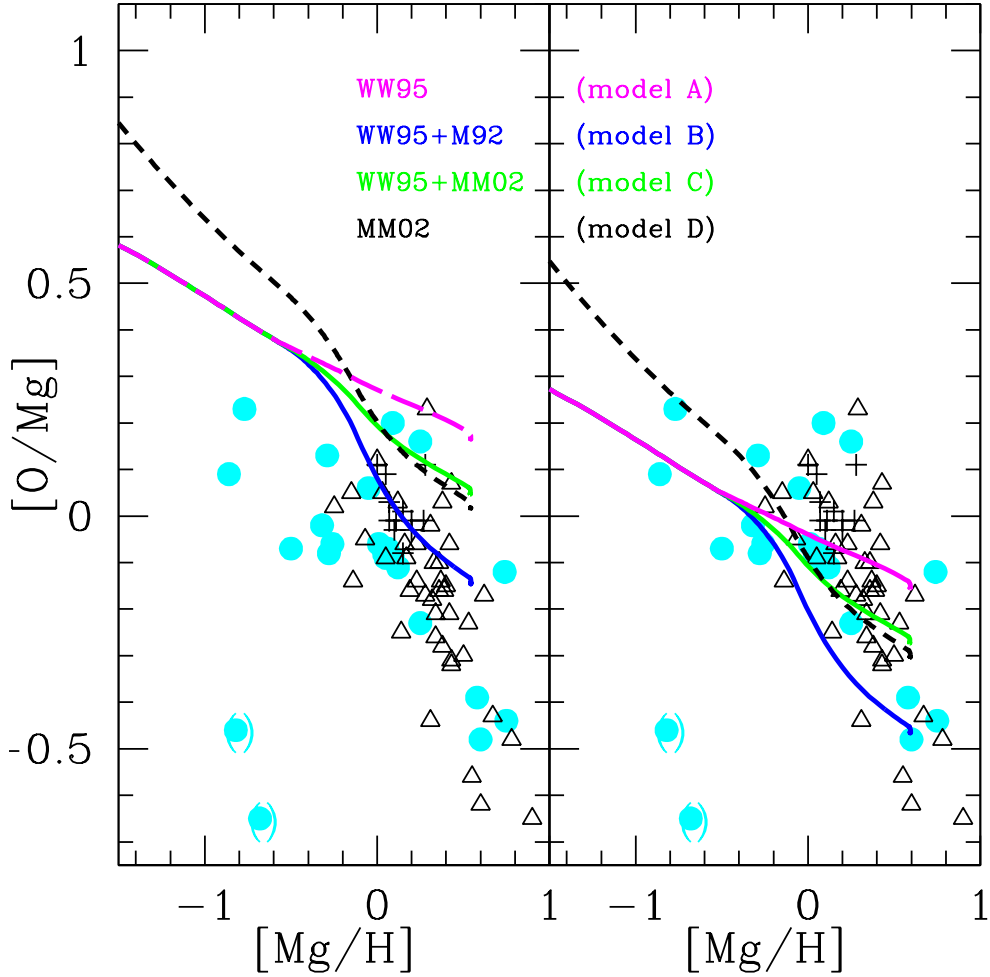}
\caption{ Comparison between the predictions of our 4 models for [O/Mg] vs [Mg/H]
and the observations in the bulge.
The observational data for the bulge are:
the filled circles are the data from MFR09 and FMR07;
the  open triangles are by Lecureur et al. (2007);
the plus signs are  the infrared results from Rich \& Origlia (2005).
Note that the two bulge stars in parenthesis (from FMR07) show the effect of proton
burning products in their atmospheres, therefore, they have probably suffered 
a reduction in the envelope oxygen abundances via stellar evolution,
so their oxygen abundances do not reflect the bulge composition. Left panel: the model results are normalized to 
Asplund et al. (2005) solar abundances. Right panel: the model results are normalized to the solar abundances 
of Grevesse \& Sauval (1998).}\label{omgmgh}
\end{center}
\end{figure}
 Finally, we should take in account that another option to explain the disagreement between 
predictions and observations is represented by the uncertainties in the observational data. \\
In Fig.~\ref{omgmgh} we show the [O/Mg] vs. [Mg/H] in the bulge. 
This figure is similar to the one in McWilliam et al. (2008) with the addition of the predictions of 
models C and D not present in that paper. Another difference with the same figure in 
McWilliam et al. (2008) is that here we adopted interpolated yields, as explained before. 
This makes the [O/Mg] ratio to decline less abruptly. We note that, apart from a zero 
point problem, the best slope for the [O/Mg] ratio for [Mg/H]$>0$ is still obtained 
with the M92 yields (model B), but models C and can still be acceptable. 
Concerning the zero point, as already discussed in our previous paper, the agreement with observations depends also on 
the assumed solar abundances: here we 
have normalized all models to the Asplund et al. (2005) abundances. If we had assumed Grevesse \& Sauval (1989) normalization, 
as in Ballero et al. (2007a), the predicted [O/Mg] would decrease by $\sim 0.2$ dex, in very good agreement with the data, 
as shown in the figure.
A better agreement could be obtained also by adopting 
a steeper than Ballero et al. (2007a) IMF for the bulge, 
as discussed for Fig.~\ref{bulgetheory}, which would  lower the [O/Mg] ratio by
$\sim$ 0.2 dex. However, a steeper IMF for the bulge would not account for the stellar metallicity distribution nor for 
the mean metallicity in the bulge (see Ballero et al. 2007b, Pipino et al. 2008). Finally, part of the problems could reside 
in the still uncertain Mg yields.

\section{Discussion and Conclusions}

We have reviewed stellar CNO abundances in the thin and thick disks and the bulge of the
Galaxy.  Because the bulge red giant stars envelopes are contaminated by first dredge-up,
CN processed, material in their envelopes, we estimate the their primordial [C/O] ratios 
from C$+$N minus a correction for the original nitrogen abundance ($\sim$0.1 dex).  
We showed that with careful measurements this method can give [C/O] ratios from red giants
identical to unmixed dwarf/turnoff stars in the thin and thick disks.

From our review of such CNO abundances in Galactic bulge red giants we find a large 
increase in the [C/O] ratio ($>$0.5 dex) that is tightly correlated with metallicity
between [O/H]=$-$0.1 and $+$0.4 dex.  This observed change in bulge [C/O] ratio
can be reproduced with bulge chemical evolution models that incorporate
yields from massive stars including the effects of metallicity-dependent stellar winds
and rotation; in particular our models would best match the observations with mass-loss
rates intermediate between the values adopted by M92 and MM02.  This supports the idea
that the decline in [O/Mg] and [O/Fe] seen in the bulge near solar metallicity is
due a decline in oxygen yields in massive stars by metallicity-dependent winds, as
suggested by McWilliam \& Rich (2004) and McWilliam et al. (2008).  This is important
because it resolves the apparent discrepancy in interpretation of the bulge formation
timescale based on simplistic interpretation of the bulge trends of [Mg/Fe] and [O/Fe].
These conclusions on the bulge [C/O] ratio support a rapid bulge formation timescale ($<$1Gyr).

To better match the slope of the bulge [C/O] ratio with [O/H] we found it necessary to
interpolate the predicted stellar yields, rather than employ a step function change in
yield at solar metallicity.  We employ the same interpolation technique for our models
of the thin disk [C/O] trend with [O/H], and we find a remarkably good correspondence between
our model results and the thin disk observations. In particular, we find that the [C/O]
slope in the thin disk is best matched with the yields of MM02.  In our models approximately
50\% of the thin disk carbon comes from low mass stars.

M08 found that the bulge and thick disk [O/Fe] versus [Fe/H] trends are identical; our
new oxygen results in MFR09 agree with this conclusion.  Remarkably, the data examined here
shows that the [C/O] versus [O/H] locus in the thick disk is also practically identical to the 
trend in the Galactic bulge.  We assume that the common trends in these two systems is due to
higher star formation rates, and faster formation timescales, than occurred in the thin
disk.  However, the higher [Mg/Fe] in the bulge than the thick disk (FMR07) is evidence that 
the bulge formed more quickly.  Still, while the bulge and thick disk
probably formed fairly rapidly it is remarkable that these two systems have such different mean
metallicities ($\Delta$[Fe/H]$\sim$ 0.5 dex).  

 As mentioned earlier, Ballero et al. (2007a,b)
employed a shallow IMF (i.e. more high mass stars) in order to fit the bulge stellar metallicity distribution as well as
the bulge mean metallicity with a rapid formation timescale.  Similarly, it may be that while the thick disk and the bulge 
both formed fairly quickly (at least relative to the thin disk),
metallicities may largely reflect different SF efficiencies (lower in the thick disk, Chiappini 2009), and/or different IMFs.
 In other words, our best model for the bulge cannot reproduce the mean metallicity and the metallicity distribution of 
the thick disk stars.

While our models naturally explain the increase in the [C/O] ratio to higher metallicity,
there is a plateau at low [O/H] in the bulge data; a similar plateau is also seen in our models,
but $\sim$0.2 dex lower than the observations.  While this plateau problem could, in principle,
be solved with a change in the bulge IMF this would lead to a conflict with the observed mean
metallicity and the rapid bulge formation timescale indicated by [Mg/Fe]; thus, we rejected
appeals to the IMF to solve the problem.  The under-predicted plateau suggests to us that there
is a missing source of carbon in our models, and/or an under-production of oxygen.  
 Extremely metal poor stars ($Z \sim 10^{-8}$) can be very fast rotators, suffer heavy mass loss and 
produce that extra C (Hirschi 2006, Chiappini \& al. 2006), but to  explain the plateau such a production 
should extend also to stars with higher metallicities. 
Another cause, although prosaic, can be an incomplete stellar evolution physics, wind physics, or
error in the adopted $^{12}$C($\alpha$,$\gamma$)$^{16}$O rate. We highlight an alternative explanation:
stripping of the envelopes of massive stars in binary systems through Roche lobe overflow, which
would work even at low metallicity where winds don't.  We presume that this form of envelope
stripping would also result in an increase in carbon yield at the expense of oxygen.
 Finally, the disagreement between theory and observations could be ascribed to observational 
uncertainties.

\section{Acknowledgments}
G.C. acknowledges financial support from the Fondazione Cassa di Risparmio di Trieste.
G.C., F.M.and C.C. acknowledge financial support from PRIN2007-MUR (Italian Ministry of University and Research). 
Prot.2007JJC53X-001.
C.C. acknowledges financial support from the Swiss National Foundation (SNF).
F.M. thanks R.M. Rich for many illuminating discussions.
We thank the referee, Martin Asplund, for his careful reading and his suggestions.

\end{document}